\documentclass{article}
\usepackage{ijcai19}

\usepackage{times}
\usepackage{soul}
\usepackage{url}
\usepackage[hidelinks]{hyperref}
\usepackage[utf8]{inputenc}
\usepackage[small]{caption}
\usepackage{graphicx}
\usepackage{amsmath}
\usepackage{booktabs}
\usepackage{verbatim}
\usepackage{paralist}

\title{Governance by Glass-Box:\\ Implementing Transparent Moral Bounds for AI Behaviour}

\author{
Andrea Aler Tubella\footnote{Contact Author}\and
Andreas Theodorou\and
Frank Dignum\And
Virginia Dignum\\
\affiliations
Ume\aa \ University\\
\emails
\{andrea.aler, andreas.theodorou, frank.dignum, virginia.dignum\}@umu.se
}

\begin{document}

\maketitle

\begin{abstract}
Artificial Intelligence (AI) applications are being used to predict and assess behaviour in multiple domains which directly affect human well-being. However, if AI is to improve people's lives, then people must be able to trust it, by being able to understand what the system is doing and why. Although transparency is often seen as the requirement in this case, 
realistically it might not always be possible, whereas the need to ensure that the system operates within set moral bounds remains.

In this paper, we present an approach to evaluate the moral bounds of an AI system based on the monitoring of its inputs and outputs. We place a `Glass-Box' around the system by mapping moral values into explicit verifiable norms that constrain inputs and outputs, in such a way that if these remain within the box we can guarantee that the system adheres to the value. The focus on inputs and outputs allows for the verification and comparison of vastly different intelligent systems; from deep neural networks to agent-based systems.

The explicit transformation of
abstract moral values into concrete norms brings great benefits in terms of explainability; stakeholders know exactly how the system is interpreting and employing relevant abstract moral human values and calibrate their trust accordingly.
Moreover, by operating at a higher level we can check the compliance of the system with different interpretations of the same value.

\end{abstract}

\section{Introduction}
Artificial Intelligence (AI) 
has the potential to greatly improve our autonomy and wellbeing, but to be able to interact with it effectively and safely, we need to be able to trust it. For example, automatic elevators were fully developed as early as in 1900. Yet, most people at the time were too uncomfortable to ride in them, citing safety concerns and support for elevator operators. It took a strike in 1945, which left New York paralysed, and a huge industry-led PR push in the early 50s to change people's minds. At the same time, the American National Standards Institute updated the \textit{Safety Code for Elevators first issuance}, later to be known as standard \textit{A17.1}, to establish minimum-safety requirements.

As this example shows, building public trust in robotics and artificial intelligence at large requires a multi-faceted approach: it is both a societal challenge and a technical one. Trust in the technology we use is a requirement for well-being and in itself requires governance mechanisms that enable us to understand ---or at least audit--- our systems to ensure compliance \cite{BrysonWinfield2017}. As such, it is becoming increasingly important to be able to verify whether intelligent systems comply with existing legal frameworks and, even further, to understand how the system is interpreting and employing relevant human values and moral principles. I.e. what does it mean for an AI system to be fair? How is fairness implemented into the system and does it satisfy the relevant legal framework? How can we compare two different systems in terms of their fairness quality?

Trust in AI is often linked to algorithmic transparency, however, transparency is more than just ensuring algorithm visibility \cite{Theodorou2017ConnScience}. By transparency we mean that the different factors that influence the decisions made by algorithms should be visible, or transparent, to the people who use, regulate, and are impacted by systems that employ those algorithms, i.e. the \textit{understandability} of a specific model \cite{lepri2018fair}. 
Transparency is also seen as the, and is often taken as, a requisite for algorithmic accountability \cite{BrysonWinfield2017}. 
However, decisions made by predictive algorithms can be opaque because of many factors, which may not always be possible or desirable to eliminate. These include technical (the algorithm may not lend itself to easy explanation), economic (the cost of providing transparency may be excessive, including the compromise of trade secrets), and social (revealing input may violate privacy expectations)\footnote{ACM statement on transparency: \url{http://www.acm.org/binaries/content/assets/public-policy/2017_usacm_statement_algorithms.pdf}} \cite{crawford2016}.
Since human decisions can also be quite opaque, as are the decisions made by corporations and organisations, mechanisms such as audits, contracts, and monitoring are in place to regulate and ensure attribution of accountability. In this paper, we investigate the possibility of applying this type of approaches with the aim to realise trustworthy AI systems that can contribute to human well-being.

Ultimately, building trustworthy AI requires a focus on three core principles: \textit{responsibility}, \textit{accountability}, and \textit{transparency} \cite{dignum2017ijcai}. Each one is necessary but not sufficient to achieve Trustworthy AI. Transparency does not imply accountability nor responsibility but complements and extends these. 
The goal of transparency is providing sufficient information to ensure at least safe usage and human accountability \cite{Bry}. For this, there is a need to at least be able to validate and verify the system for auditing purposes, a \textit{sine qua non} to ensure compliance with any legal or ethical principles a system is required (or meant to) follow. 
However, validation and verification procedures are highly dependant on the legislation and ethical framework that the system is deployed on, and depend on the specific contextual interpretations that have been employed to ground abstract principles (e.g. fairness or privacy) into concrete system functionalities.

In this paper, we describe the Glass-Box approach, a two-stage methodology which takes into account the contextual interpretations of abstract principles taking a Design for Values perspective \cite{Hoven05,vdpoel2013} and allows for the validation of systems by circulating around the challenges presented by black boxes. By mapping moral values into explicit verifiable norms that constrain and direct inputs and outputs, we place a `Glass-Box' around the system in such a way that if these remain within the box we may say that the system adheres to the value in a specific context. To produce such a mapping it is necessary in a first interpretation stage to make explicit design decisions concerning the meaning of values in different legislative and ethical frameworks,  subsuming them into progressively more fine-grained norms and finally into specific functionalities concerning the inputs and outputs of the system. 
The second stage consists of observing the behaviour of the system (by means of formal verification, simulation, or monitoring) and checking for its adherence to the low-level requirements on inputs and outputs specified in the interpretation stage. Depending on which of these requirements are met, we are able to say in which context the system can be considered to adhere to a value.

The explicitness in the interpretation of values allows stakeholders to know exactly what it means for a system to adhere to a certain value, and thus to calibrate their trust accordingly. Furthermore, the system can include its adherence to a value as an explanation for its actions \cite{Pizza2017}. Design for Values is made easier, as it is a matter of implementing the right Glass-Box with the appropriate norms. The focus on inputs and outputs allows for the verification and comparison of vastly different intelligent systems, from neural networks to agent based systems. Moreover, the versatility of our approach allows us to check the compliance of the system with different interpretations of the same value. This compliance checking is not only considered a minimum-level transparency for a deployed system, but also necessary to ensure the due diligence of a system and, therefore, enforce attribution of legal or ethical accountability if needed \cite{BrysonWinfield2017}.

First, we discuss our motivation and need to Design for Values. Next, we present to the reader our Glass-Box approach. In the penultimate section we provide an example of how our approach can be applied to a real-world intelligent system. We conclude the paper with a discussion and conclusions of the work presented.

\section{Responsible Artificial Intelligence}\label{sec:responsible}

In all areas of application, where AI is applied to make decisions that affect people and society, the most important issue to consider is perhaps the need to rethink responsibility \cite{dignum2017ijcai}. 
Responsible Artificial Intelligence is about human responsibility for the development of intelligent systems along fundamental human principles and values, to ensure human flourishing and well-being in a sustainable world. That is, AI reasoning should be able to take into account societal values, moral and ethical considerations; weigh the respective priorities of values held by different stakeholders in various multicultural contexts; explain its reasoning; and guarantee transparency. Responsible AI is more than the ticking of some ethical `boxes' in a report, or the development of some add-on features, or switch-off buttons in AI systems. Rather, responsibility is fundamental to autonomy and should be one of the core stances underlying AI research. Ethical and social requirements for responsible AI can be classified into three areas \cite{Dignum2018}:
\begin{itemize}
	\item \textbf{Ethics \underline{by} Design}: the technical/algorithmic integration of ethical reasoning capabilities as part of the behaviour of artificial autonomous system;
	\item \textbf{Ethics \underline{in} Design}: the regulatory and engineering methods that support the analysis and evaluation of the ethical implications of AI systems as these integrate or replace traditional social structures;
	\item \textbf{Ethics \underline{for} Design}: the codes of conduct, standards and certification processes that ensure the  integrity of developers and users as they research, design, construct, employ and manage artificial intelligent systems.
\end{itemize}
Many organisations and nations have produced, or are in the process of announcing, statements on the values or principles that should guide the development and deployment of AI in society\footnote{These include but are not limited to: the United Nations, the European Union, UK's Engineering and Physical Sciences Research Council, ACM, and IEEE Standards Association.}. The current emphasis on the delivery of high-level statements on AI ethics may also bring with it the risk of implicitly setting the `moral background' for conversation about ethics and technology \cite{greene2019better} as being about abstract principles. Often these statements lack precise normative frameworks that can enable the understanding of how ethical values are interpreted and implemented in concrete applications. Moreover, there is not a one-to-one link between values and actions/functionalities. In general, every action will contribute to more than one value, either positively or negatively, while systems need to balance between different values.

Ensuring socially beneficial outcomes of AI relies on resolving the tension between incorporating the benefits and mitigating the potential harms of AI. In short, simultaneously avoiding the misuse and underuse of these technologies \cite{ai4people2018}. 
Needless to say that compliance with the law is always a fundamental requirement of any system, but it is also significantly insufficient. In fact, compliance with the law is the minimum that should be ensured but an ethical approach goes further than legal compliance  \cite{floridi2018soft}.

Values and principles are dependent on the socio-cultural context~\cite{turiel2002culture}; they are often only implicit in deliberation processes. 
Theories and methods are needed that elicit and integrate societal, legal and moral values into all stages of development (analysis, design, construction, deployment and evaluation) of any intelligent system. 
Design for Values methodologies~\cite{friedman2006,Hoven05} provide rational procedures for designing artefacts under the guidance of moral values. 
During the development of AI systems, taking a Design for Values approach means that the process needs to include explicit activities for~\cite{aldewereld2014}:
\begin{inparaenum}[(i)]
	\item the identification of societal values, 
	\item deciding on a moral deliberation approach (e.g. through algorithms, user control or regulation), and
	\item link values to formal system requirements  and concrete functionalities \cite{Aldew}.
\end{inparaenum}

For example, consider the development of an intelligent recruitment application. A value that can be assumed for this system is \textit{fairness}. However, fairness can have different normative interpretations, such as \textit{equal access to resources}, or \textit{equal opportunities}, which can lead to very different actions. It is, therefore, necessary to make explicit which interpretation(s) is taken into the design. This decision may be informed by domain requirements and regulations, e.g. a choice for \textit{equal opportunities} meets the legal requirement established in national law. Finally, we still need to be explicit about how this norm is implemented in the system, which depends on the context but is also influenced by the personal views and cultural background of those deciding on the interpretation. Machine learning literature identifies different functional interpretations of the equal opportunities view of fairness, e.g. demographic parity~\cite{Pedreschi2008} and equal odds~\cite{Dwork2012} amongst others. The Design for Values approach makes these choices explicit and supports formal verification. 

\begin{figure*}[ht]
\centering
\includegraphics[width=0.7\textwidth]{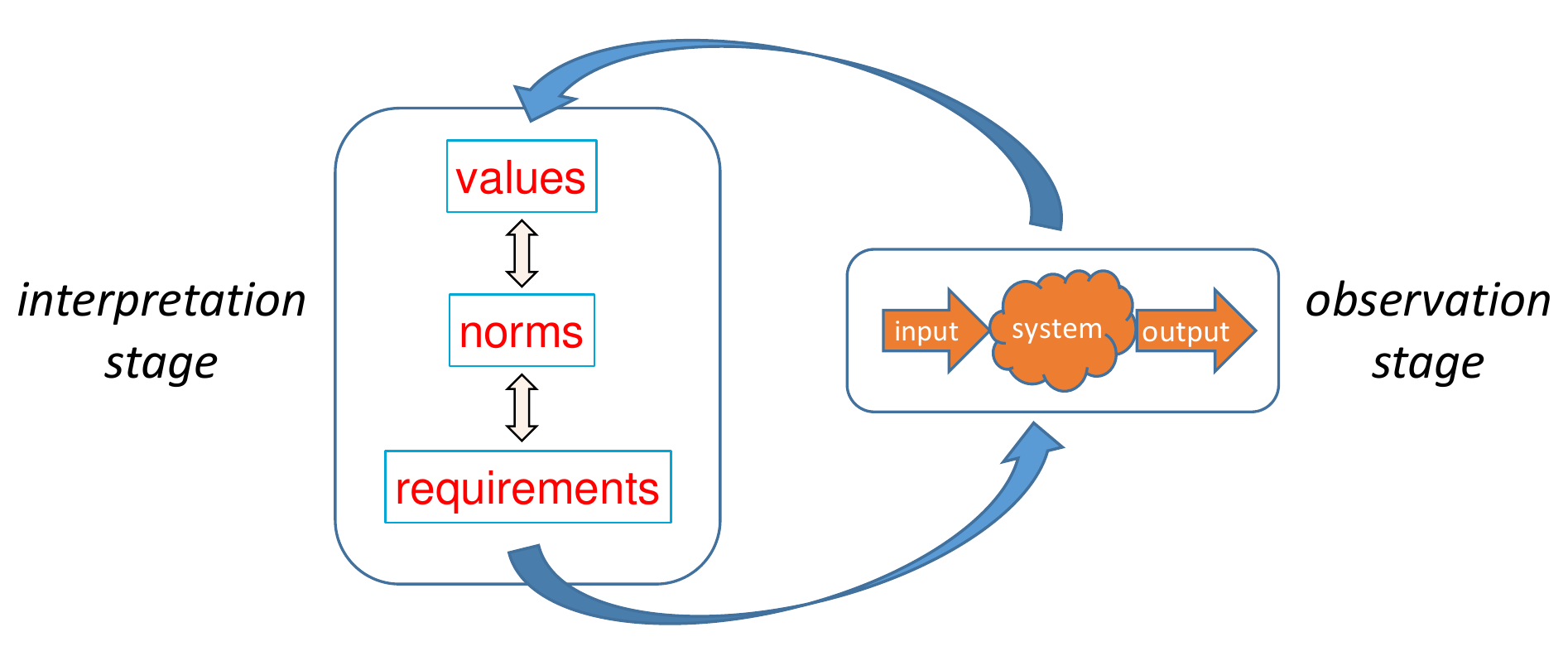}
\caption{The two stages of the Glass-Box Approach: an Interpretation stage, where values are translated into design requirements, and an Observation stage, where we can observe and qualify the behaviour of the system.}
\label{fig:gb}
\end{figure*}

\section{Glass-Box Approach}
In this section, we provide a sketch of the Glass-Box approach, as depicted in Figure \ref{fig:gb}. The approach includes two phases:
\begin{itemize} 
    \item Interpretation stage: the explicit and structured process of translating values into design requirements. This is a bi-directional relation, where a top-down \textit{counts-as} specification relation describes the translations of higher level elements into lower level, contextualised, concepts; and a \textit{for-the-sake-of} relation connects lower level elements, such design requirements to higher level elements, such as more general norms and values \cite{vdpoel2013}. Interpretation decisions define the way systems are ultimately built. In this sense, they can be seen as Searle's constituitive rules \cite{Searle1995TheReality}.
    \item Observation stage: based on the functionalities specified in the Interpretation phase, we can observe and qualify the behaviour of the AI application. In this stage compliance with the values is approximated to the evaluation of the system with respect to the requirements specified in the interpretation stage. Different methods can be used including formal verification, simulation or monitoring. 
\end{itemize}

In the the rest of this section, we describe these stages in more detail.

\subsection{Interpretation Stage}

The first step of the interpretation stage entails a translation from abstract values into concrete norms comprehensive enough so that fulfilling the norm will be considered as adhering to the value. Following a Design for Values approach (see section \ref{sec:responsible}), a shift from abstract to concrete necessarily involves careful consideration of context. Normative systems are often described in deontic-based languages, which allow for the representation of obligations, permissions and prohibitions. In the Glass-Box Approach however, we aim to not only describe the norms themselves, but also the exact contextual connection between abstract and concrete concepts.

Several authors have proposed \textit{counts-as} statements as a means to formalise interpretations \cite{Aldew}.  With this relation, we can build logical statements of the form: ``X counts as Y in context C" \cite{Searle1995TheReality,Jones1995APower}. \textit{Counts-as} is a contextual subsumption relation that describes the translations of higher level elements into lower level, contextualised, concepts.

The \textit{counts-as} operator admits formalisations based in modal logic \cite{Jones1995APower,Grossi2006Counts-as:Logic} and description logic \cite{Grossi2006OntologicalInstitutions} and thus it lends itself to implementations \cite{Aldew}. Furthermore, the interactions of \textit{counts-as} relations operating in different contexts may be formalised as well \cite{Grossi2007DesigningSystems}. With this approach we can formally represent the explicit relations between abstract and concrete concepts, and, given that the relations between concepts are dependent on the context in which that relation is evaluated, the definition of the context of those relations is made explicit as well. In this formalisation, the conjunctions and disjunctions of different norms will therefore stand for the explicit interpretation of a value in a specific, explicit context.

The second step in the interpretation stage is the concretisation of norms into specific system requirements. In the Glass-Box approach these requirements will be given in terms of the inputs and outputs of the intelligent system. The connection of these lower level requirements to higher level elements can be described in terms of  a \textit{for-the-sake-of} relation \cite{vdpoel2013}. This relation would allow for the formalisation of statements of the form ``X is done for the sake of Y''. 

As a simple example which we expand in the next section, consider a system advising on the attribution of mortgages. We wish for the value \textit{privacy} to be implemented in the system, taking into account different perspectives. At the end of the interpretation stage, after discussion with stakeholders, we may arrive at a norm postulating that from the point of view of the customer, revealing only the personal information available to the tax agency counts as respecting the value of \textit{privacy}. Then, this norm can be distilled into a functionality by ensuring that only this type of data is entered into the system. Other perspectives and values are then similarly concretised and translated into functionalities.
 
At the end of the interpretation stage we will therefore have built an abstract-to-concrete hierarchy of norms where the highest level is made-up of values and the lowest level is composed of fine-grained concrete requirements for the intelligent system only related to its inputs and outputs. The intermediate levels are composed of progressively more abstract norms, and the connections between nodes on each level are contextual. The concrete requirements inform the observation stage of our approach, as they indicate what must be verified and checked. On the other hand, this hierarchy can be used after the observation stage to provide high-level transparency for a deployed system: depending on which requirements are being fulfilled, we can provide explanations for how and exactly in which context the system adheres to a value.

\subsection{Observation Stage}
As result of the previous stage, the list of requirements that provides an interpretation of the desirable ethical principles has been identified. In this stage, the behaviour of the system is evaluated with respect to the values by studying its compliance with these requirements. 
In \cite{Vazquez-Salceda2007FromBehavior} two requirements for norms to be enforceable are identified:  verifiability i.e. the low-level norms must allow for being machine-verified given the time and resources needed, and computational tractability, i.e. whether the functionalities comply with the norms can be checked on any moment in a fast, low cost way. Note that this is a requirement for the observation stage and not necessarily for the design stage! Hence, some of the norms chosen for the design stage might be easily implementable, but hard to monitor. For example, a neural network approach to implement a mortgage decision, in which the neural net is trained on all decisions of the last years can provide an implementation not to deviate from decisions in similar cases. However, it is not easy to monitor or govern that the decisions never deviate more than a certain percentage from similar cases.

In order to ensure that the Glass-Box approach is enforceable, governance mechanisms that include the specification of quality of service levels. As an example that we will elaborate in the next section, a minimum `fairness-quality' level for a mortgage-decision system, may be to guarantee that e.g. anyone who requests a mortgage of less than 30\% of his/her liquid loan should be granted it. More refined quality levels include the guarantee that all similar applicants receive the same treatment. Observing these levels poses different constraints to the Glass-Box framework: whereas for the former only the behaviour with respect to one given applicant is needed, the latter is dependent of data about many applicants within a given time frame and region.

The mechanisms to observe this behaviour can be implemented without knowledge about the internal workings of the system under observation, by monitoring input and output streams. We insist on this feature as we do not always have access to the internals of the system, neither do we always have access to the designs of a system. 
Either way, the main challenge is the computational tractability of these checks on the design requirements and their implementation, and the fact that systems, and values and norms, evolve. Part of the challenge is then on determining the required granularity of the Glass-Box: a too rough approximation will possibly cap many potentially compliant behaviours, whereas a too specific Glass-Box may limit the adaptation of the AI system. 

Given that we have designed the system using the value-based design method sketched in this section it is possible to use the norms from the design stage to derive suitable norms for the observation stage. However, if these design stage norms are not available we have to design them based on the general requirements of the system taking into account a balance between enforceability and effectivity.
Still, another issue comes up in this stage, which is that we might want the system to comply to given external regulations, such as the regulation on fair treatment of customers in the national law.

\section{Use Case}

\begin{figure*}[ht]
    \centering
    \includegraphics[scale=0.7]{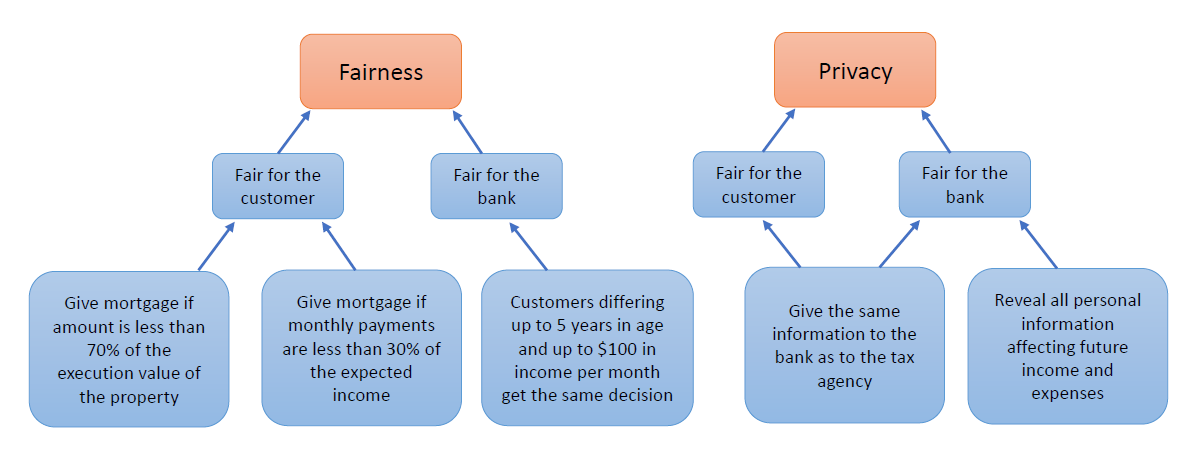}
    \caption{Hierarchy of norms and values identified at the interpretation stage. Blue arrows represent \textit{counts-as} relationships.}
    \label{fig:hierarchy}
\end{figure*}

Let us consider a system that decides on mortgages as an example. The starting point is to check what \textit{fair} means in the context of mortgage decisions. The general idea of fairness is that costs and benefits are divided in a way that is acceptable for all stakeholders. In this example there are at least three ways to view this. The decision should be fair between the bank and the customer; e.g. the bank should not have to take a high risk of losing the money and the customer should get a mortgage if she can afford it. The decision should also be fair between customers; e.g. two customers in comparable situations should get the same decision and if their situations are different, then it should be clear how this influences the decision. Finally, the decision should also remain fair over time; for example, a bank should not drastically change its policy from one day to another without prior warning. 

For each of the perspectives of fairness above we can check which norms might support the value of fairness. For the bank a norm could be to give a mortgage if the amount of the mortgage is less than 70\% of the execution value of the property. In that case the bank can still sell the property and after deducing costs and still recover all of its money if the customer fails to pay. Another norm would be that if the monthly payments of the mortgage are less than 30\% of the expected income of the customer the mortgage is affordable and thus should be awarded. Many more norms could of course be given to define in more detail what a bank is considering to be a risk and what a customer is considering to be affordable. We do not argue that these exact norms are the best ones to use, but rather that with our approach any norms that we choose are explicitly coupled to a perspective of fairness.

Similarly, we can create norms to guarantee fairness between customers. In this perspective, norms will define what it means that customers are in a similar situation and should get a similar decision. For example, it can be decided that customers in the same 5 year age bracket and incomes not differing more than \$100 per month should get the same decision on loans that do not differ more than \$5000. One can easily see that here is the point where biases come in: indeed, the criteria that are used to determine whether customers are in similar situations can greatly bias the decision. Do we only check the income of the customer for the next year or do we want an expected income for the next 10 years? The latter would favor tenured staff as temporary staff cannot guarantee an income after two years or so. Again, we do not claim that this is the only interpretation of fairness for this scenario, but rather we want to illustrate that for each value there are multiple perspectives and their importance also differs depending on the stakeholder.

In order to make `better' decisions on the loan, the bank can argue that it should look at more factors than just current income; e.g. the personal life of the customer. However, that may violate the \textit{privacy} of the customer ---another value that needs to be respected. What if at the time of the application a couple has enough income to get a loan according to the norms, but they are expecting a child and they are planning to switch to part-time work after the baby is born? Such a move could reduce their income to just 75\% of their current one. One could argue that it would be `fair' not to withhold this information from the bank at the time of the loan application. However, one can also argue that this is private information and that many things can occur before the baby is born and, thus, they should not be obliged to tell the bank anything about these plans.
Another norm could be to oblige the customers to give all information to the bank that they also give to their country's tax agency, but nothing more. This could be a balanced compromise between revealing private information and providing sufficient financial information. Again, how far this solution is the best possible norm is subject to debate. The important point here is that this norm is constituted exactly for the purpose of balancing between these two values in this context. If customers would prefer not to give that much information they might find another bank that gives more priority to customer privacy, but then is more conservative in giving loans as the uncertainty ---and, hence, the risk--- is higher. 
Thus, different values can steer behaviour in different directions. After the norm identification process, we obtain a hierarchy of norms and values illustrated in Figure \ref{fig:hierarchy}

The last step of interpretation is from norms to functionalities. In our example this means that the bank could employ some AI system to calculate the possible decisions on mortgage requests. This system should comply with the norms, but can still have considerable room to optimise the decisions. For example, it can still give customers a loan even if they get a monthly instalment higher than 30\% of their income: the norm only forbids refusing on the basis of this criteria, but if the customer has a very high income it might be that the income left after paying the monthly instalment is still very high and enough to warrant the risk. The decisions presented here are a kind of rule based system, we could also consider a neural network that is trained on existing customer data and the constraints given by the norms. Furthermore, a decision on a mortgage is not just a yes/no decision, but can vary in the interest requested on the loan. 

The above indicates how values can be implemented in a system. The verification of compliance is a separate, but related, case. Given the issues raised in the previous section on norm verification, we need to consider the impact of the outcome of the system on the users, the dynamics of the context, the continuity of the output values as related to the input parameters and the efficiency and tractability of the verification.

If the bank uses the 30\% norm it is easy to check and thus cost effective, efficient and will be fair between customers and over time. However, if more fine grained fairness is required and not only the decision between getting a mortgage or not, but also the interest rate should be fair, the scenario becomes far more complex. In this case the Glass-Box is fitted quite tight around the implemented system and thus needs to be formulated in a way to give room for flexibility and adaptiveness of the system, but at the same time guarantee the value of fairness. A Glass-Box defined such that an interest rate on a mortgage can never be more than 1,5*central bank interest may work well in ''usual" circumstances, but is very difficult to maintain if the interest of the central bank goes to zero. On the other hand, a Glass-Box with an interest rate that is calculated by simply adding 2 to the interest of the central bank might be too loose and not guarantee much fairness. A different formulation where the fairness of the decisions is taken more literally and each decision is compared to all other decisions the bank has made until that point has its own flaws: it becomes very inefficient (considering that one has to first determine `similarity' between customers and their situation).
From the above it can be seen that by using the different layers of modelling it is possible to guarantee the compliance to certain values (and priorities between values).

\section{Discussion and Conclusions}

Achieving trustworthy AI systems is a multifaceted complex process, which requires both technical and socio-legal initiatives and solutions to ensure that we always align an intelligent system's goals with human values. Core values, as well as the processes used for value elicitation, must be made explicit and that all stakeholders are involved in this process. Furthermore, the methods used for the elicitation processes and the decisions of who is involved in the value identification process are clearly identified and documented. Similarly, all design decisions and options must also be explicitly reported; linking system features to the social norms and values that motivate or are affected by them. This not unlike to how software developers in regulated industries already have to map new features and bug reports to code changes for ISO9001 accreditation \cite{ISO9001}. This should always be done in ways that provide inspection capabilities ---and, hence, traceability--- for code and data sources to ensure that data provenance is open and fair.

The focus on inputs and outputs allows for the verification and comparison of vastly different intelligent systems, from neural networks to agent based systems. Moreover, the versatility of our approach allows us to check the compliance of the system against different interpretations of the same value, e.g. the American interpretation of `fair use' for data handling is different compare to the one set by the EU. Furthermore, the system can include its adherence to a value as an explanation for its actions, providing a high-level transparency necessary to ensure the due diligence of a system.

The Glass-Box approach opens interesting avenues for follow-up research. The development of a formalism to express concrete input/output requirements would be an interesting first step. Furthermore, such a formalism could be linked to formal verification procedures.
In addition, although the mechanisms for describing the  connections between requirements and norms and between norms and values are outlined in this paper, further insight may be gained through the development of an all encompassing requirement-to-value formalism that would allow for the description of the full hierarchy of concepts. On the other hand, given a Glass-Box, it would be interesting to study whether the systems that fulfil its requirements can be characterised. Last, technical implementations of concrete Glass-Boxes for a system may be developed.

While we strive to make tools and technologies like our Glass-Box approach widely available and accepted, we must ensure legal paths to address by ownership and/or usage our responsibility and accountability. Otherwise, we may have the same problem as the one we often have with private-owned militias; lack of effective responsibility and accountability. Governance mechanisms, such as standards and legislation to enforce technical solutions like the Glass-Box, can ensure that any moral responsibility or legal accountability is properly appropriated by the relevant stakeholders, together with the processes that support redressing, mitigation and evaluation of potential harm, and means to monitor and intervene on the system's operation. Finally, responsible AI ---and any policy at large---- also requires informed participation of all stakeholders, which means that education plays an important role, both to ensure that knowledge of the potential impact of AI is widespread, as well as to make people aware that they can participate in shaping societal development.

\section*{Acknowledgements}

This work was partially supported by the Wallenberg AI, Autonomous Systems and Software Program (WASP) funded by the Knut and Alice Wallenberg Foundation, and by the European Union’s Horizon 2020 research and innovation programme under grant agreement No 825619.

\bibliographystyle{named}
\bibliography{ijcai19}
\end{document}